\documentclass[12pt]{article}
\usepackage{amssymb,axodraw,graphicx}
\def\beq{\begin{equation}}
\def\eeq{\end{equation}}
\def\beqa{\begin{eqnarray}}
\def\eeqa{\end{eqnarray}}

\begin{document}

\renewcommand{\thefootnote}{\fnsymbol{footnote}}
\setcounter{footnote}{1}

\mbox{ } \\[-1cm]
\mbox{ }\hfill TUM--HEP--518/03\\
\mbox{ }\hfill hep--ph/0306303\\
\mbox{ }\hfill \today\\

\begin{center}
  {\Large\bf Decay of Super-Heavy particles:\\ User guide of the SHdecay program} \\[2mm]
\bigskip
C. Barbot\footnote{barbot@ph.tum.de}

\end{center}

\bigskip

\begin{center}
\it Physik Dept., TU M\"{u}nchen, James Franck Str., D--85748
    Garching, Germany \\[1mm]
\end{center}

\bigskip
\bigskip
\bigskip

\begin{abstract}

\vskip 0.5cm

\noindent

I give here a detailed user guide for the C++ program SHdecay, which
has been developed for computing the final spectra of stable particles
(protons, photons, LSPs, electrons, neutrinos of the three species and
their antiparticles) arising from the decay of a super-heavy $X$
particle. It allows to compute in great detail the complete decay
cascade for any given decay mode into particles of the Minimal
Supersymmetric Standard Model (MSSM).  In particular, it takes into
account all interactions of the MSSM during the perturbative cascade
(including not only QCD, but also the electroweak and 3rd generation
Yukawa interactions), and includes a detailed treatment of the SUSY
decay cascade (for a given set of parameters) and of the
non-perturbative hadronization process. All these features allow us to
ensure energy conservation over the whole cascade up to a
numerical accuracy of a few per mille.  Yet, this program also allows
to restrict the computation to QCD or SUSY-QCD frameworks. I detail
the input and output files, describe the role of each part of the
program, and include some advice for using it best.

\end{abstract}

\newpage

{\large \bf PROGRAM SUMMARY}
\vspace{1cm}

{\it Title of program: } SHdecay

{\it Computer and operating system: } Program tested on PC running 
Linux KDE and Suse 8.1

{\it Programming language used: } C with STL C++ library and using the
standard gnu {\tt g++} compiler.

{\it No. Lines in distributed program: } around 7400 lines.


{\it Nature of Physical Problem: } obtaining the energy spectra of the
final stable decay products (protons, photons, electrons, the three
species of neutrinos and the LSPs) of a decaying super-heavy $X$
particle, within the framework of the Minimal Supersymmetric Standard
Model (MSSM). It can be done numerically by solving the full set of
DGLAP equations in the MSSM for the perturbative evolution of the
fragmentation functions $D_{p_1}^{p_2}(x,Q²)$ of any particle $p_1$
into any other $p_2$ ($x$ is the energy fraction carried by the
particle $p_2$ and $Q$ its virtuality), and by treating properly the
different decay cascades of all unstable particles and the final
hadronization of quarks and gluons. In order to obtain proper results
at very low values of $x$ (up to $x \sim 10^{-13}$), NLO color
coherence effects have been included by using the Modified Leading Log
Approximation (MLLA).

{\it Method of solution: } The DGLAP equations are solved by a four
order Runge-Kutta method with a fixed step.

{\it Typical running time: } around 35 hours for the first run, but
the most time consuming sub-programs can be run only once for most
applications. 

\newpage
{\large \bf LONG WRITE-UP}

\section{Introduction}
\label{sec:introduction}

Although they obviously have never been observed, very different types
of super-heavy (SH) particles (with masses up to the grand unification
scale, at $10^{16}$ GeV and even beyond) are predicted in a number of
theoretical models, e.g. grand unified and string models. But even
without invocating these particular theories, their existence is quite
natural; indeed, it is known that the Standard Model of particle
physics (SM) cannot be the fundamental theory\footnote{That is the
  reason why we will work here within the framework of supersymmetric
  theories, and more specifically within the Minimal Supersymmetric
  Standard Model (MSSM), which offers a very promising extension of
  the SM at energies up to the grand unification scale $M_{GUT} \sim
  10^{16}$ GeV, where a remarkable unification of couplings naturally
  occurs. For a review, see e.g. \cite{reviewMartin}.}, but only an
effective theory at low energy (say, up to the TeV region); thus one
should find one (or more) fundamental energy scale at higher energies,
and there is reason to believe that some (super-heavy) particle(s)
would be associated to this new scale.

If such particles exist, they should have been produced in large
quantities during the first phases of the universe, especially during
or immediately after inflation \cite{creat}. Their decay could have
had a strong influence on the particle production in the early
universe; this is certainly true for the decay of the inflatons
themselves. Moreover, the decay of such particles has been proposed as
an alternative solution of the ultra-high energy cosmic ray (UHECR)
problem.  Indeed, if such particles have survived until our
epoch\footnote{At first sight, this assumption seems to be rather
  extreme, but many propositions have been made in the literature for
  explaining such a long lifetime; for example, the $X$ particles could
  be protected from decay by some unknown symmetry, which would only
  be broken by non-renormalizable operators of high orders occurring
  in the Lagrangian ; or they could be ``trapped" into very stable
  objects called topological defects (TDs), and released when the TDs
  happen to radiate. For a review, see \cite{reviewSigl}.}, their decay
could explain the existence of particles carrying energy up to
$10^{20}$ eV, which have been observed in different cosmic ray
experiments over the past 30 years \cite{expts} and still remain one of
the greatest mysteries in astrophysics. For studying in detail the
predictions of these models, a new code taking into account the full
complexity of the decay cascade of super-heavy particles is needed.

The program SHdecay has been designed for this purpose. It allows to
compute the spectra of the final stable decay products of any $N$-body
decay of a super-heavy $X$ particle, independently of the model
describing the nature of this particle; the only fundamental
assumption behind this work is that such a particle will decay into
``known'' particles of the MSSM\footnote{This is a very reasonable
  assumption when one considers the success of the MSSM in accounting
  for the unification of couplings at $E \sim 10^{16}$ GeV that we
  mentioned before. In order not to destroy this beautiful feature,
  one assumes in the so-called ``desert hypothesis'' that there is no
  ``new physics'', thus no new energy scale, between the TeV region
  and the GUT scale. In this case, the only available particle content
  is the one of the MSSM.}.

In this article, which is supposed to be a user guide for the public
code SHdecay, I first give a brief overview of the physics involved in
the decay of a SH particle (section~\ref{sec:physics}). I then give a
more technical presentation of the problem (section~\ref{sec:theory}),
where I describe the DGLAP evolution equations \cite{AP} which have to
be solved (section~\ref{subsec:DGLAP}) and our implementation of the
MLLA approximation in the low $x$ region (section~\ref{subsec:MLLA}).
I then turn to the SHdecay program itself; I describe in
section~\ref{sec:black_box} the ``master program'' contained in the
package, which partly allows to use the whole program as a ``black
box''. In section~\ref{sec:programs}, I present the organigram of the
code and describe all its components in detail; I also list all the
options of the master program. The definition of our ``compound
particles'' and the id's used to label them are given in Appendix A.
Finally, an example of typical SHdecay input/output files in the QCD
case are given in section ``{\bf Test Run Input and Output}'', as well
as a plot of two final output fragmentation functions obtained in the
most general case (MSSM framework).

\section{Physics background}
\label{sec:physics}

I first briefly describe the physical steps involved in the decay
cascade of a super-heavy $X$ particle in the framework of the MSSM, as
they are illustrated on fig 1, before turning to the relevant
technical details in the next section. A full description of the
physics of $X$ decays is given in refs~\cite{bd1,bd2}.

As I explained in the introduction, the basic assumption is that the
$X$ particle decays into $N$ particles of the MSSM (in fig 1 a pair of
squark/antisquark). These particles will initially have a time-like
virtuality ${\cal O}(M_X)$, hence each of them initiates a parton
cascade, following the known physics at lower energy. As can be seen
in fig 1, the first part of the cascade is a perturbative
fragmentation process involving all couplings of the MSSM\footnote{In
  contrast to previous works
  \cite{Berezinsky:2000,Coriano:2001,Sarkar:2001,ToldraLSP}, we
  considered in our treatment all gauge interactions as well as third
  generation Yukawa interactions, rather than only SUSY--QCD ; note
  that at energies above $10^{20}$ eV all gauge interactions are of
  comparable strength. The inclusion of electroweak gauge interactions
  in the shower gives rise to a significant flux of very energetic
  photons and leptons, which had not been identified in earlier
  studies.}. It will continue until the virtuality has decreased
enough, to a scale where both SUSY and $SU(2) \otimes U(1)$ will break
(for simplicity we are considering a unique SUSY mass scale $M_{SUSY}
\sim 1$ TeV for all sparticles\footnote{We call ``superparticles'' or
  simply ``sparticles'' all superpartners of the SM particles which
  are defined in the MSSM.}, which is symbolized by the first vertical
dashed line in fig 1.) All the on-shell massive sparticles produced at
this stage will then decay into Standard Model (SM) particles and the
only (possibly) stable sparticle, the so-called Lightest
Supersymmetric Particle (LSP)\footnote{The stability of the LSP is
  taken for granted in the MSSM, as a consequence of the introduction
  of a new unbroken discrete symmetry called R-parity, which prevents
  dangerous violations of baryon and lepton number conservation ; in
  most of the parameter space, the LSP is the particle designed by
  $\tilde{\chi}_1^0$, the lightest neutralino.}. The heavy SM
particles, like the top quarks and the massive bosons, will decay too,
but the lighter quarks and gluons will continue a perturbative
partonic shower until they have reached either their on-shell mass
scale or the typical scale of hadronization ($Q_{\rm had} \sim 1$
GeV), the second vertical dashed line of fig 1.  At this stage, the
color force becomes too strong and the partons cannot propagate freely
anymore, being forced to combine into colorless hadronic states.
Finally, the unstable hadrons and leptons will also decay, and only
stable particles will remain and propagate (through the intergalactic
space, in case of UHECRs), namely the protons, photons, electrons, the
three species of neutrinos and the LSP (and their antiparticles).

The perturbative part of the shower is treated through the numerical
solution of DGLAP evolution equations \cite{AP} extended to the
complete spectrum of the MSSM. These equations describe the evolution
with virtuality of the so-called ``fragmentation functions'' (FFs),
which are describing the fragmentation of any fundamental particle $i$
into any other $j$; the FFs, generically called $D_i^j(x,Q)$, depend
on the virtuality $Q$ and on the energy fraction $x = 2E/M_X$ (where E
is the energy of the new particle $j$). The evolution equations
include the running of the associated coupling constants. We worked
out all the FFs of the MSSM by solving these equations for all fields
(in the interaction basis) between $M_{SUSY}$ and $M_X$. At the
breaking scale $M_{SUSY}$ we applied the canonical unitary
transformation to these FFs in order to obtain those of the mass
eigenstates, and computed the decay cascade of the supersymmetric part
of the spectrum.  I used here the public code ISASUSY \cite{Isasusy}
(which is a subpart of the ISAJET code) to describe the allowed decays
and their branching ratios, for a given set of SUSY parameters. If
R-parity is conserved, the LSP remains stable, and we obtain its final
spectrum at this step ; the rest of the available energy is
distributed between the SM particles.  After another perturbative
cascade down to $Q_{\rm had} \sim \max(m_{\rm quark},1 {\rm GeV})$, as
stated before, the quarks and gluons will hadronize. I used the
results of \cite{Poetter}, which are fitted to available data, as
input functions for describing the hadronization and convoluted them
with our previous results for the FFs of quarks and gluons (according
to the factorization theorem of QCD, see for example \cite{QCDrev}).
During the complete cascade, we paid special attention to the
conservation of energy.  We are able to follow the energy conservation
through the complete evolution up to a few per mille.

The physical meaning of the FF $D_X^i (x,M_X)$ can be directly
understood as the number of $i$ particles carrying an energy $x \times
M_X$ produced through the decay of one X particle. Combined with a
particular distribution of $X$ particles (and with possible
propagation effects), it allows to compute the flux of any final
stable particle at any energy `$\leq M_X$. Especially, this analysis
can be applied to a distribution of $X$ particles in the universe, in
order to compute the expected flux of Ultra High Energy Cosmic Rays
for a given top-down model. Some applications of this code have
already been presented in \cite{bdhh1,bdhh2}.

\section{Theory of a super-heavy $X$ particle decay}
\label{sec:theory}

\subsection{Calculation of the fragmentation functions}
\label{subsec:DGLAP}

Consider the two--body decay of an ultra--massive $X$ particle of mass
$m_X \gg 10^{20}$ eV into a $q\bar{q}$ pair, in the framework of the
MSSM. This triggers a parton shower, which can be understood as
follows. The $q$ and $\bar q$ are created with initial virtuality $Q_X
\sim \frac{M_X}{2}$. Note that $Q_X^2 > 0$. The initial particles can
thus reduce their virtuality, i.e. move closer to being on--shell
(real particles), by radiating additional particles, which have
initial virtualities $< Q_X$. These secondaries then in turn initiate
their own showers. The average virtuality and energy of particles in
the shower decreases with time, while their number increases (keeping
the total energy fixed, of course). As long as the virtuality $Q$ is
larger than the electroweak or SUSY mass scale $M_{\rm SUSY}$, all
MSSM particles can be considered to be massless, i.e. they are all
active in the shower. However, once the virtuality reaches the weak
energy scale, heavy particles can no longer be produced in the shower;
the ones that have already been produced will decay into SM particles
plus the lightest superparticle (LSP), which we assume to be a stable
neutralino. Moreover, unlike at high virtualities, at scales below
$M_{\rm SUSY}$ the electroweak interactions are much weaker than the
strong ones; hence we switch to a pure QCD parton shower at this
scale. At virtuality around 1 GeV, nonperturbative processes cut off
the shower evolution, and all partons hadronize. Most of the resulting
hadrons, as well as the heavy $\tau$ and $\mu$ leptons, will
eventually decay. The end product of $X$ decay is thus a very large
number of stable particles: protons, electrons, photons, the three
types of neutrinos and LSPs; I define the FF into a given particle to
include the FF into the antiparticle as well, i.e. I do not
distinguish between protons and antiprotons etc.

Note that at most one out of these many particles will be observed on
Earth in a given experiment. This means that we cannot possibly
measure any correlations between different particles in the shower;
the only measurable quantities are the energy spectra of the final
stable particles, $d \Gamma_X / d E_P$, where $P$ labels the stable
particle we are interested in.\footnote{Of course, this just describes
  the spectrum of particle $P$ at the place where $X$ decays. In all
  cosmological applications, the code SHdecay will have to be related
  to a ``propagation code'' taking into account all the by
  interactions with the interstellar and intergalactic medium (see
  e.g. \cite{reviewSigl}). I will not address this issue in this
  paper.} This is a well--known problem in QCD, where parton showers
were first studied. The resulting spectrum can be written in the form
\cite{QCDrev}
\begin{equation} \label{def_ff}
\frac {d \Gamma_X} {d x_P} = \sum_I \frac {d \Gamma(X \rightarrow I)}
{d x_I} \otimes D^P_I(\frac{x_P}{x_I}, Q_X^2),
\end{equation}
where $I$ labels the MSSM particles into which $X$ can decay, and we
have introduced the scaled energy variable $x = 2 E / m_X$; for a
two--body decay, $d \Gamma(X \rightarrow I) / d x_I \propto
\delta(1-x_I)$. The convolution is defined by $f(z) \otimes g(x/z) =
\int_{x}^{1} f(z)g\left(\frac{x}{z}\right)\,\frac{dz}{z}$.

All the nontrivial physics is now contained in the fragmentation
functions (FFs) $D^P_I(z,Q^2)$. Roughly speaking, they encode the
probability for a particle $P$ to originate from the shower initiated
by another particle $I$, where the latter has been produced with
initial virtuality $Q$. This implies the ``boundary condition''
\begin{equation} 
\label{boundary}
D^J_I(z,m_J^2) = \delta_I^J \cdot \delta(1-z),
\end{equation}
which simply says that an on--shell particle cannot participate in the
shower any more. For reasons that will become clear shortly, at this
stage we have to include all MSSM particles $J$ in the list of
``fragmentation products''. The evolution of the FF with increasing
virtuality is described by the well--known DGLAP equations
\cite{QCDrev}:
\begin{equation} 
\label{dglap}
\frac{dD_I^J} {d\ln(Q^2)} (x,Q) =
\sum_K \frac{\alpha_{KI} (Q^2)} {2\pi} P_{KI}(z) \otimes
D_K^J(x/z,Q^2),
\end{equation}
where $\alpha_{KI}$ is the coupling between particles $I$ and $K$, and
the splitting functions $P_{KI}$ describes the probability for
particle $K$ to have been radiated from particle $I$. As noted
earlier, for $Q > M_{\rm SUSY} \sim 1$ TeV we allow all MSSM particles
to participate in the shower. Since we ignore first and second
generation Yukawa couplings, we treat the first and second generations
symmetrically. $I,J,K$ in eq.(\ref{dglap}) thus run over 30 particles:
6 quarks $q_L, u_R, d_R, t_L, t_R, b_R$, 4 leptons $l_L, e_R, \tau_L,
\tau_R$, 3 gauge bosons $B, W, g$, the two Higgs fields of the MSSM,
and all their superpartners. Here we sum over all color and $SU(2)$
indices (i.e., we assume unbroken $SU(2)$ symmetry), and we ignore
violation of the CP symmetry, so that we can treat the antiparticles
exactly as the particles. All splitting functions can be derived from
those listed for SUSY--QCD in \cite{Jones}; explicit expressions will
be given in a later paper. The starting point of this part of the
calculation is eq.(\ref{boundary}) at $Q = M_{\rm SUSY}$. This leads
to $30 \times 30$ FFs $D_I^J$, which describe the shower evolution
from $Q_X$ to $M_{\rm SUSY}$.

At scales $Q < M_{\rm SUSY}$ all interactions except those from QCD
can be ignored. Indeed, at scales $Q < Q_0 \simeq$ 1 GeV QCD
interactions become too strong to be treated perturbatively. leading
to confinement of partons into hadrons. This nonperturbative physics
cannot be computed yet from first principles; it is simply
parameterized, by imposing boundary conditions on the
$D_i^h(z,Q_0^2)$, where $h$ stands for a long--lived hadron and $i$
for a light parton (quark or gluon). Here I used the results of
\cite{Poetter}, where the FFs of partons into protons, neutrons, pions
and kaons are parameterized in the form $Nx^{\alpha}(1-x)^{\beta}$,
using fits to LEP results. Note that these functions are only valid
down to $x = 0.1$; for smaller $x$, mass effects become relevant at
LEP energies. However, at our energy scale these effects are still
completely irrelevant, even at $x=10^{-7}$. I chose a rather simple
extrapolation at small $x$ of the functions given in \cite{Poetter},
of the form $N'x^{\alpha'}$. I computed the coefficients $N'$ and
$\alpha'$ by imposing energy conservation. Starting from these
modified input distributions, and evolving up to $Q = M_{\rm SUSY}$
using the pure QCD version of eq.(\ref{dglap}), leads to FFs
$D_i^h(x,M^2_{\rm SUSY})$ which describe the QCD evolution (both
perturbative and non--perturbative) at $Q < M_{\rm SUSY}$.

Finally, the two calculations have to be matched together. First we
note that ``switching on'' $SU(2)$ and SUSY breaking implies that we
have to switch from weak interaction eigenstates to mass eigenstates.
This is described by unitary transformations of the form $D^S_I =
\sum_J |c_{SJ}|^2 D^J_I$, with $\sum_S |c_{SJ}|^2 = \sum_J |c_{SJ}|^2
= 1$; here $S$ stands for a physical particle. I used the ISASUSY
code \cite{Isasusy} to compute the SUSY mass spectrum and the
$|c_{SJ}|$ corresponding to a given set of SUSY parameters. The decay
of all massive particles $S$ into light particles $i$ is then
described by adding $\sum_S D^S_I \otimes P_{iS}$ to the FFs $D^i_I$
of light particles $i$. I assumed that the $x-$dependence of the
functions $P_{iS}$ originates entirely from phase space. In this
fashion each massive particle $S$ is distributed over massless
particles $i$, with weight given by the appropriate $S \rightarrow i$
decay branching ratio. Note that this step often needs to be iterated,
since heavy superparticles often do not decay directly into the LSP, so
that the LSP is only produced in the second, third or even fourth
step. All information required to model these cascade decays have
again been taken from ISASUSY. The effects of the pure QCD shower
evolution can now be included by one more convolution, $D^h_I(Q_X) =
\sum_i D^h_i(M_{\rm SUSY}) \otimes D^i_I(Q_X)$. Finally, the decay of
long--lived but unstable particles $\mu, \tau, n, \pi, K$ has to be
treated; this is done in complete analogy with the decays of particles
with mass near $M_{\rm SUSY}$.

\subsection{Coherence effects at small $x$: the MLLA solution}
\label{subsec:MLLA}

So far we have used a simple power law extrapolation of the hadronic
(non--perturbative) FFs at small $x$. This was necessary since the
original input FFs of ref.\cite{Poetter} are valid only for $x \geq
0.1$. As noted earlier, we expect our treatment to give a reasonable
description at least for a range of $x$ below 0.1. However, at very
small $x$, color coherence effects should become important
\cite{Basics_of_QCD}. These lead to a flattening of the FFs, giving a
plateau in $x D(x)$ at $x_{\rm plateau} \sim \sqrt{Q_{\rm had}/M_X}
\sim 10^{-8}$ for $M_X = 10^{16}$ GeV. One occasionally needs the FFs
at such very small $x$. For example, the neutrino flux from $X$ decays
begins to dominate the atmospheric neutrino background at $E \sim
10^5$ GeV \cite{neutrinos, bdhh1}, corresponding to $x \sim 10^{-11}$
for our standard choice $M_X \sim 10^{16}$ GeV. In this subsection we
therefore describe a simple method to model color coherence effects in
our FFs.

This is done with the help of the so--called limiting spectrum derived
in the modified leading log approximation. The key difference to the
usual leading log approximation described by the DGLAP equations is
that QCD branching processes are ordered not towards smaller
virtualities of the particles in the shower, but towards smaller
emission angles of the emitted gluons; note that gluon radiation off
gluons is the by far most common radiation process in a QCD shower.
This angular ordering is due to color coherence, which in the
conventional scheme begins to make itself felt only in NLO (where the
emission of two gluons in one step is treated explicitly). It changes
the kinematics of the parton shower significantly. In particular, the
requirement that emitted gluons still have sufficient energy to form
hadrons strongly affects the FFs at small $x$. For sufficiently high
initial shower scale and sufficiently small $x$ the MLLA evolution
equations can be solved explicitly in terms of a one--dimensional
integral \cite{Basics_of_QCD}. This essentially yields the modified FF
describing the perturbative gluon to gluon fragmentation $D_g^g$. In
order to make contact with experiment, one makes the additional
assumption that the FFs into hadrons coincide with $D_g^g$, up
to an unknown constant; this goes under the name of ``local
parton--hadron duality'' (LPHD) \cite{LPHD}. Here I use the fit of
this ``limiting spectrum'' in terms of a distorted Gaussian
\cite{distorted_gaussian}, which (curiously enough) seems to describe
LEP data on hadronic FFs somewhat better than the ``exact'' MLLA
prediction does. It is given by
\beq 
\label{gaussian}
F_i(\xi,\tau) \equiv xD_i(x,Q) =
\frac {\bar{n}_i} {\sigma \sqrt{2\pi} } \exp {\left[ \frac {1} {8} k +
\frac{1}{2} s \delta - \frac{1}{4} (2+k) \delta^2 + \frac{1}{6} s
\delta^3 + \frac{1}{24} k \delta^4 \right]},
\eeq
where $\bar{n}_i$ is the average multiplicity. The other quantities
appearing in eq.(\ref{gaussian}) are defined as follows:
\beqa 
\label{gauss_coef}
\tau &=& \log{\frac{Q}{\Lambda}} \,,\nonumber\\
\xi &=& \log{\frac{1}{x}} \,,\nonumber\\
\bar{\xi} &=& \frac{1}{2} \tau \left( 1 + \frac {\rho} {24}
\sqrt{\frac{48}{\beta \tau} } \right) + {\cal O}(1)\,,\nonumber\\
\sigma &=& \langle (\xi - \bar{\xi})^2 \rangle^{1/2} =
\sqrt{ \frac{1}{3} } \left( \frac{\beta}{48} \right)^{1/4} \tau^{3/4}
\left( 1 - \frac{1}{64} \sqrt{ \frac{48\beta} {\tau} } \right) + {\cal
O} (\tau^{-1/4}) \,,\nonumber\\
\delta &=& \frac{ \xi -\bar{\xi}} {\sigma} \,,\nonumber\\
s &=& \frac{ \langle (\xi - \bar{\xi})^3 \rangle }{\sigma^3} =
-\frac {\rho}{16} \sqrt{ \frac{3} {\tau} } \left( \frac {48}
{\beta\tau} \right)^{1/4} + {\cal O}(\tau^{-5/4}) \,,\nonumber\\
k &=& \frac{ \langle (\xi - \bar{\xi})^4 \rangle }{\sigma^4} =
- \frac {27} {5\tau} \left( \sqrt{ \frac{1}{48} \beta \tau } - \frac
{1}{24} \beta \right) + {\cal O}(\tau^{-3/2}) \,.
\eeqa
where $\beta$ is the coefficient in the one--loop beta--function of
QCD and $\rho = 11 + 2 N_f/27$, $N_f$ being the number of active
flavors. Eqs.(\ref{gaussian}) and (\ref{gauss_coef}) have been derived
in the SM, where $\beta = 11 - 2 N_f/3$. Following
ref.\cite{limiting_spectrum} I assume that it remains valid in the
MSSM, with $\beta = 3$ above the SUSY threshold $M_{\rm SUSY}$ and
$\rho = 11 + 8/9$. 

When comparing MLLA predictions with experiments, the overall
normalization $\bar n_i$ (which depends on energy) is usually taken
from data. We cannot follow this approach here, since no data with $Q
\sim M_X$ are available. Moreover, usually MLLA predictions are
compared with inclusive spectra of all (charged) particles. We need
separate predictions for various kinds of hadrons, and are therefore
forced to make the assumption that all these FFs have the same
$x-$dependence at small $x$. This is perhaps not so unreasonable; we
saw above that the DGLAP evolution predicts such a universal
$x-$dependence at small $x$. I then match these analytic solutions
(\ref{gaussian}), (\ref{gauss_coef}) with the hadronic FFs $D_I^h$ we
obtained from DGLAP evolution and our input FFs at values $x_0^h$,
where for each hadron species $h$ the matching point $x_0^h$ and the
normalization $\bar n_h$ are chosen such that the FF and its first
derivative are continuous; One typically finds $x_0 \sim 10^{-4}$. Note
that this matching no longer allows to respect energy conservation
exactly. However, since the MLLA solution begins to deviate from the
original FFs only at $x \sim 10^{-7}$, the additional ``energy
losses'' are negligible.

\section{How to use SHdecay as a black box}
\label{sec:black_box}

\setcounter{footnote}{1}

Here I would like to describe how to use this program as easily as
possible, ignoring the different internal components, and considering
the whole program as a ``black box''. I just want to stress that the
price to pay is running time... Indeed, certain component programs of
this code are pretty time consuming - especially the first one
(DGLAP\_MSSM), which is solving a set of 30 integro-differential
equations over orders of magnitude in virtuality, and needs around 30
hours of running on a modern computer\footnote{For processors of 1 GHz
  and above, the running time seems to be almost independent of the
  exact frequency, and there is no gain of time with increasing
  frequencies.}.  Yet, in most applications, DGLAP\_MSSM and its
``brother'' DGLAP\_QCD have to be run only once.  Moreover,
DGLAP\_MSSM can be ``cut'' into smaller pieces which can be run
independently on different computers. This will require more detailed
knowledge of this program; see sec~\ref{sec:programs}.

There is another point I want to insist on : although SHdecay is a
self-contained code, it requires two Input files that have to be
obtained from an other program, like the public code ISASUSY : these
two files contain all information about the SUSY spectrum (masses and
mixing angles), and the decay modes of the sparticles, top quark and
Higgses, with the associated branching ratios (BRs). In order to keep
the completeness of the furnished code, I implemented a personalised
version of ISASUSY\footnote{I used the version 7.43 of ISASUSY} in
this package in a fully transparent way for the user. Nevertheless, if
you want to use another available code giving the same information, or
even an updated version of ISASUSY, you will have to work by yourself
for obtaining the two output files (called by default ``Mixing.dat''
and ``Decay.dat'', and stored in the Isasusy directory) in the
required format. I will come back to this point in
section~\ref{sec:programs}.

\subsection{Installation of SHdecay}

SHdecay has been written for a UNIX or Linux operating system. It
certainly can be used on a computer using windows with a C++ compiler,
but in that case you won't be able to use the provided makefiles. In
the following, I am describing the procedure for using SHdecay on a
UNIX/Linux plateform.  

Once you have downloaded the compressed package ``SHdecay.tar.gz'',
just write the file into an empty directory, and decompress it with
the command :\newline

{\bf tar -xzf SHdecay.tar.gz\newline}

\noindent
It will create a directory SHdecay and install inside all files and
subdirectories you need. A list of all files contained in the package
with a brief description is providing in the file ``Listing.txt''.
Then enter the directory SHdecay and compile the ``master program'' by
typing:\newline

{\bf run\_SHdecay\newline}

\noindent
You can now call the ``master program'':\newline

{\bf SHdecay.exe\newline}

\clearpage

\noindent
The following menu should appear:\newline
\bigskip
\bigskip

 *************************** SHdecay.c  ***************************

    0 : Compile all.

\bigskip

    1 : Run all programs.

    2 : Run all programs but Isasusy.

\bigskip

    3 : DGLAP\_MSSM     (DGLAP evolution for the FFs between M\_SUSY and M\_X).

    4 : Isasusy         (MSSM spectrum and decay modes).
 
    5 : Susy1TeV        (SUSY and SU(2)*U(1) breaking; SUSY decay cascade).

    6 : DGLAP\_QCD      (Pure QCD DGLAP evolution down to Q\_had).

    7 : Fragment\_maker (Non-perturbative FFs at Q\_had).

    8 : Less1GeV        (Hadronization and SM decays).

    9 : Xdecay          (Final FFs for a given X decay mode).

\bigskip
 *******************************************************************\newline
\bigskip

You first have to compile all programs with the option ``0''. You'll
certainly get a few warnings that you can ignore (They arise from the
fact that you are compiling the program for the first time, and thus
it doesn't need to erase old files before compiling). Once it has been
done, call again the master program SHdecay.exe: now you can choose
the program you do want to run.

The 1st option allows to run the different programs as a whole black
box for given input files. You need two of them: 
\begin{itemize}
\item[a)] The first one is called ``Input.dat'' by default (but you
  may write your own with a different name: you will be asked for the
  name of this file at the very beginning of the run): it contains
  physical and technical parameters necessary for the run, as well as
  the name of output directories in which you will store the results.
  A default version of ``Input.dat'' is included (option ``/''), and
  all parameters have default values inside the program itself (option
  ``*''). Yet, you will need to write your own input data file in most
  cases. I present all the required input parameters in the next
  section.
\item[b)] The second input file is called ``SUSY.dat'' by default (but
  again, you can give it another name ; you will be asked for it
  during the run): it contains all SUSY parameters that will be
  needed in ISASUSY, as well as the names of the two output files
  mentioned above (by default ``Mixing.dat'' and ``Decay.dat'').
\end{itemize}

This option will run successively all programs contained in SHdecay,
following the organigram given in fig. 2. In this case, the user has
nothing to do but filling the two input files; yet, the required
running time will be around 36 hours on a recent computer (see the
note above) in the MSSM framework.

N.B.: The second option is the same as the first one, up to the fact
that it doesn't run the Isasusy program. It still requires the two
files ``Mixing.dat'' and ``Decay.dat'' to be present in the Isasusy
directory - no matter how you produced them. It only becomes usefull
in the case you want to free yourself from the program ISASUSY.

The next options allow to run each code individually; for a
description, see the corresponding subsection in
section~\ref{sec:programs}.

\subsection{Parameters of the ``Input.dat'' file.}

The two first parameters concern two options of the program :
\begin{itemize}
\item[a)] ``Theory'' is an integer describing the theoretical
  framework in which the computation will be done. Four options are
  available : 1 for Minimal Supersymmetric Standard Model (MSSM), 2
  for Standard Model (SM), 3 for SUSY-QCD, and 4 for QCD alone. This
  option concerns the particles which will be included in the
  perturbative cascade at high energy, while solving the set of DGLAP
  equations. As a result, it also determines in which primary
  particles the $X$ particle is allowed to decay. But it doesn't
  affect the decays of particles with masses $\leq M_{SUSY}$ at all,
  which are governed by known physics and require the full SM
  spectrum. For example, in the QCD framework (Theory = 4), only
  quarks and gluons will be taken into account in the DGLAP equations,
  which means that we will neglect all but QCD couplings ; yet the top
  quark will still decay into $b W$ and $W$ in leptonic as well as quark
  channels at $M_{SUSY}$ !). By default I use the MSSM framework
  (Theory = 1).\newline Caution : for the moment, the SM option is not
  fully implemented.
\item[b)] ``MLLA'' is an integer describing the approximation made at
  low $x$ for the non-perturbative hadronic FFs. Two choices are
  possible :
\begin{itemize} 
\item[0 :] one assumes a power law extrapolation at low $x$ for the
  final FFs (which indeed have a power law shape at the end of the non
  perturbative cascade). But this extrapolation doesn't take into
  account the saturation of the FFs at low $x$ due to the appearence
  of mass and color effects.
\item[1 :] one assumes the Modified Leading Log Approximation (MLLA)
  \cite{Basics_of_QCD} with the implementation of a distorted gaussian
  at small x \cite{distorted_gaussian}, in order to take the color
  effects into account. For details on the treatment, I refer to
  \cite{bd2}. 
\end{itemize}
By default the MLLA approximation is used (MLLA = 1).
\end{itemize}

The next set of parameters describes the physical inputs of the program (all
masses, energies and virtualities are given in GeV):
\begin{itemize}
\item[a)] ``Nb\_output\_virtualities'' gives the number of
  different values for the $X$ mass you want to study.  By default two
  final masses are stored (Nb\_output\_virtualities = 2).\newline
  Caution : for each $X$ mass you will get a lot of output files, for
  example $30*30=900$ in the MSSM framework !
\item[b)] In ``Output\_virtualities\_DGLAP\_MSSM(GeV)'', you should
  specify the exact values of the virtualities at which you want to
  store the output FFs. (Exactly as many values as you asked for in
  ``Nb\_output\_virtualities'' !). By default two final virtualities
  (i.e. $X$ masses !) which are stored are: $10^{12}$ and $10^{16}$
  GeV\footnote{Caution: As the name of this parameter indicates,
    this option {\it only concerns} the outputs of the first program
    DGLAP\_MSSM. The following programs will treat only {\it one}
    case, the one required by parameter $M_X$. See the technical
    section for the reasons of this choice.}.
\item[c)] ``$M_X$'' (in GeV) doesn't give exactly the {\it mass} of the X
    particle\footnote{In fact, the {\it real mass} of the $X$ particle will be
      $2*M_X$.}, but the {\it initial virtuality} of the decay
    products of X, that is, the highest virtuality of the perturbative
    cascade ; of course, it must be one of the values given in
    ``Output\_virtualities\_DGLAP\_MSSM(GeV)'', at which the output
    FFs have been stored. This parameter will be used by all other
    programs following DGLAP\_MSSM. If you want to do the complete
    treatment for different $M_X$ masses, you will have to run these
    other programs as many times as necessary, with the different
    values of $M_X$. By default $M_X$ is set to $10^{16}$ GeV (the GUT
    scale).
\item[d)] ``N-body\_X\_decay'' must contain the value of $N$ for a N-body
  decay mode of the $X$ particle. By default I consider a 2-body decay.
\item[e)] ``X\_decay\_mode'' contains the details of the $N$-body decay
  mode you want to study. Of course, it must contain as many particles
  as asked in ``N-body\_X\_decay'' ; the id's of the different particles
  are given in Appendix A of this manual. By default $X$ is
  decaying into two SU(2) doublet first/second generation left
  ``(anti)quarks'' $q_L$, with id 1.
\item[f)] The ``$M_{\rm SUSY}$'' parameter must contain the virtuality (in
  GeV) at which both SUSY and $SU(2) \otimes U(1)$ are broken ; it is
  also the virtuality at which all sparticles (but the LSP), top
  quarks and heavy bosons decay. By default $M_{\rm SUSY} = 1000$ GeV.
\item[g)] $Q_{\rm had}$ gives the virtuality at which
  hadronization of the lightest quarks and gluons occurs and the
  non-perturbative fragmentation functions are convoluted with the
  perturbative ones. By default $Q_{\rm had} = 1$ GeV.
\end{itemize}

{\bf Important note} : for practical reasons, it is only possible to choose
{\it powers of ten} for all energy scales. Fortunately, such
restriction is not too handicaping, because the DGLAP evolution
equations are only {\it logarithmic} in energy.

The next set of input parameters (namely $X_{\rm Size}$, $X_{\rm extra
  Size}$, Part\_init, Part\_fin and $X_{\rm min}$) are essentially
technical, and I recommend to keep the default values, which have been
carefully adjusted in order to maximize the precision and minimize the
time needed for running. They will be described in more detail in the
technical sections.

The next set of parameters only has a practical purpose : give their
names to the output file and directories where the results will be
stored (The output themselves will be described in the next
subsection) :
\begin{itemize}
\item[1)] ``Region'' is a suffix which will be added to the name of
  certain output files. It could be used to label the set of SUSY
  parameters which has been used.
\item[2)] ``Output\_file'' gives the path and the name of the final
  output file where all the parameters of the run will be stored, and some
  results on the final energy carried by each type of stable particles
  will be given.
\item[3)] The names of the 5 next parameters are hopefully explicit
  enough : they describe the names of the directories where the output
  data files (containing the description of the FFs) of the different
  programs will be stored. If these directories don't exist already,
  they will be created automatically. For technical reasons, I don't
  allow to give a different output directory for each of the programs
  involved in the computation. For a first use of SHdecay as a whole,
  I advise to put all the FFs in the same directory, as it is done by
  default (``LowBeta'' being the common output directory for
  DGLAP\_MSSM, Susy1TeV, Less1GeV and X\_decay).\newline Caution : the
  output directory of Fragment\_maker contains the non-perturbative
  input FFs at low energy, which are essentially built from the
  results of \cite{Poetter}. I advise to use a special directory for
  that purpose (by default : Fragment), considering these FFs to be an
  {\it input} of SHdecay, which can be taken from another source if
  newer results become available. A similar reason lead us to choose a
  different default output directory for the program DGLAP\_QCD, which
  only depends on parameters $M_{SUSY}$ and $Q_{had}$, and thus need to be
  run only once, independently of the others, for most applications.
\end{itemize}

\subsection{Parameters of the ``SUSY.dat'' file.}

CAUTION : even if you don't want to use the ISASUSY code, you have to
fill this input file, at least with the value of $\tan \beta$
(required by the program DGLAP\_MSSM\footnote{Note that, except for
  $\tan \beta$ which determines the strength of the Yukawa couplings,
  the other SUSY parameters are not needed in SHdecay itself, but only
  in the ISASUSY program mentioned above, which provides the basic
  information about the SUSY decay cascade.}) and the two last
parameters giving the names of the two input files. These files have
to be placed in the Isasusy directory of SHdecay.

All SUSY parameters should be self explanatory. All masses should be
given in GeV. By default, the masses as well as the $\mu$ mass
parameter are chosen to be all 1000 GeV, and the trilinear couplings
are set to 1000. The optional values for the 2nd generation sfermions,
the gaugino and the gravitino masses, are set to $10^{20}$
GeV\footnote{Of course, that is not a physical value, but an internal
  convention for Isasusy.}. I refer to the user guide of
Isasusy\cite{Isasusy} for further information.

$\tan \beta$ is the usual parameter of SUSY theories defined by $\tan
\beta = \frac{<H_2^0>}{<H_1^0>}$, where the $<H_i^0>$ describe the
vacuum expectation values of the two Higgs fields of the MSSM. By
default, $\tan \beta = 10$.

\subsection{Output files}

One inconvenience of this program is that it has to store a lot of
data files, most of them being partial results which will be needed
for the next steps of the computation. For example, the program
DGLAP\_MSSM will have to store the FFs of any (s)particle of the MSSM
into any other; this requires $30 \times 30 = 900$ files for each set
of parameters\footnote{As described in \cite{bd1}, it has been assumed
  that certain MSSM particles can be treated symmetrically; this
  reduces the number of independent particles from 50 to $\sim 30$.}.
These partial results are certainly not relevant for the user who
wants to use SHdecay as a black box. So I will only describe here the
final results which are produced; all partial results will be
described in the section dedicated to the corresponding program.

In fact, there are only 2 or 3 types of relevant results :
\begin{itemize}
\item[1.] The final FFs themselves $D_i^j(x,M_X)$, of any initial
  decay product $i$ of the $X$ particle (among the 30 available
  ``particles'' of the MSSM, see Appendix A for details) into one of
  the seven stable particle $j$ (proton, photon, electron, the three
  types of neutrinos and the LSP). They are computed at the end of the
  program called Less1GeV, and stored in the corresponding output
  directory in 30 different files (all seven FFs for a given initial
  decay product $i$ of $X$ are grouped into one file); these files are
  called generically ``fragment\_i.all\_Region'', where $i$ is the
  initial decay product of $X$ defined above, and Region is the suffix
  labelling the set of SUSY parameters which has been used. Each of
  them contains 8 columns, giving respectively : the $x$ values (in
  decreasing order from 1. to X\_min), and the seven FFs into protons,
  $\gamma$, LSPs, $e^{-}$, $\nu_e$, $\nu_\mu$, $\nu_\tau$ respectively
  (more precisely the results correspond to the {\it sum} of the FFs
  for final particles {\it and} antiparticles).
\item[2.] For the user who wants to study a {\it precise decay mode} of the
  $X$ particle into $N$ particles of the MSSM, the relevant results will
  be given by the program Xdecay, and stored in the {\it same output
    directory} as the one given in Input.dat for Less1GeV. The name of
  the output file is ``frag\_X\_a\_b\_c(...).all'', where
  a,b,c,... are the id's of the $N$ corresponding particles (given in
  Input.dat as ``X\_decay\_mode'' ; the correspondence between particles
  and id's ibf s given in Appendix A). The 8 columns of the file are
  exactly the same as the ones decribed above.
\item[3.] The user might be interested in the amount of
  energy stored in each type of final stable particles. This
  information, as well as all the input values for the corresponding
  run of the program, is stored in an output file whose name is given
  as ``output\_file'' in Input.dat.
\end{itemize}
  
For convenience, in addition to the output files themselves, I provide
two functions called ``fragment\_fct'' (one in C++ and the other in
fortran 77) which allow to use any of the FFs computed in SHdecay in
another code. These functions are reading the specified input file and
computing the necessary cubic spline of the function. They are stored
in the ``Tools'' directory (``fragment\_fct.c'' for the C++ version,
and ``fragment\_fct.f'' for the fortran 77 one).

\begin{itemize}
\item[-] In C++, you will need the string type, that you can just
  include by adding:

{\bf \#include $<$string$>$}

\noindent
You then have to declare the function through the following line:

{\bf extern double fragment\_fct(double x, char* path\_file, string p\_fin);}

\noindent
You finally can call this function through the command:

{\bf fragment\_fct(x,path\_file,p\_fin)}

\noindent
Then, if ``program.c'' is the name of your program (written in the SHdecay directory), just compile it with

{\bf g++ program.c ./Tools/fragment\_fct.c ./Tools/my\_spline.c}

\item[-] In fortran 77, you just call the function with the same command:

{\bf FRAGMENT\_FCT(x,path\_file,p\_fin)}

\noindent
and compile your ``program.f'' program with:

{\bf f77 program.f fragment\_fct.f spline.f}

\end{itemize}

where 

\begin{itemize}
\item[-] $x$ is the (real) value at which one wants to compute the FF,
\item[-] path\_file is a chain of characters giving the complete
  (relative) path to the file containing the data (it must be given
  between two cotes),
\item[-] p\_fin is the final particle one is interested in (to be
  chosen between ``p'' for protons, ``gam'' for photons, ``LSP'' for
  LSPs, ``e'' for electrons, ``nu\_e'', ``nu\_mu'', or ``nu\_tau'' for
  the three species of neutrinos, or possibly ``'' if the chosen file
  only contains a single FF, as it is the case for all the partial
  results of the code)\footnote{It must be clear that p\_fin is just
    needed in order to distinguish between the 7 FFs which are stored
    in the same output file when coming from Less1GeV or Xdecay. For
    other files it must be set to ``''.}. It also must be given
  between two cotes.
\end{itemize}


\section{Description of the different programs}
\label{sec:programs}

\setcounter{footnote}{1}

There are mainly four successive programs treating the different parts of
the decay cascade; in order of decreasing virtuality, these are :
DGLAP\_MSSM, Susy1TeV, DGLAP\_QCD, and Less1GeV. Eventually, a
last small program called Xdecay can be run to study a particular
decay mode of the $X$ particle. I will describe in detail the role of
each of these programs, and the parameters of Input.dat they are sensitive
to. Fig~\ref{Organigram} gives a detailed organigram of the whole
code, which shows the interdependences between the different programs
and their input parameters.

The whole code SHdecay needs at two different steps the results of
other independent codes, namely :
\begin{itemize}
\item[1)] the SUSY mass spectrum, mixing angles, and decay modes of 
sparticles (with their branching ratios), all given by ISASUSY (a
subset of the Isajet code, written in Fortran 77).
\item[2)] the non-perturbative input fragmentation functions, 
computed (once and for all) from the results of \cite{Poetter} through a
program called Fragment\_maker (which is furnished). 
\end{itemize}
We'll describe these two secondary procedures in the corresponding
subsections.

Of course, the values of all the parameters written in Input.dat
should be kept the same for the four programs running successively.

All these programs have been written in C using a few C++
tools\footnote{This code is {\it not} written in an object oriented
  way !}. The compiling option of SHdecay is using the g++ compiler of
gnu (given by default on Unix and Linux OS).

I first describe all technical parameters before going into the
details of each program.

\subsection{Technical parameters}
\label{subsec:technical_param}

\begin{itemize}
\item[1)] ``$X_{\rm Size}$'' gives the number of $x$ values used to
  store the FFs on the interval [$10^{-7}$:$1-10^{-7}$]. Because of 1)
  the behaviour of the splitting functions at small x, 2) the fact
  that we are beginning with ``delta functions''\footnote{modelised
    numerically by sharp gaussians centered at 1. and normalised to
    unity between 0 and 1.} at large $x$, and 3) the definition of the
  convolution which is relating the low and large $x$ regions, the two
  extremities of our interval have to be modelled symmetrically with
  great accuracy, if we want the integration and (cubic spline)
  extrapolation procedures to be able to give results at the desired
  precision of $\sim 10^{-3}$. For this purpose I used a
  bi-logarithmic scale between [$10^{-7}$:0.5] and [0.5:$1-10^{-7}$],
  increasing the number of $x$ values towards the two extremities. By
  default, $X_{\rm Size} = 101$\footnote{Note that $X_{\rm Size}$ {\it
      has to be odd}!}, i.e. 50 $x$ values on each side of the central
  value at $x = 0.5$. Note that a smaller value could lead to false
  results, while increasing $X_{\rm Size}$ is increasing greatly the
  running time needed by all programs. So I really advise the user not
  to change this value. Note finally that the smallest $x$ value
  $10^{-7}$ has been choosen at the limit of the validity of the
  (leading order) DGLAP equations, before MLLA effects become strong
  (which happens at $\sqrt{\frac{Q_{had}}{M_X}} \sim 10^{-8}$ for $M_X
  \sim 10^{25}$ eV and $Q_{had} \sim 1$ GeV; see \cite{bd2}). At low
  $x$, the standard LO DGLAP equations will predict a power law
  behaviour\footnote{The power law can of course be extrapolated
    easily towards lower $x$, avoiding the extremely time consuming
    running of DGLAP\_MSSM on a larger $x$ interval !} (option MLLA =
  0), but the MLLA approximation (option MLLA = 1), will allow to
  parameterize some NLO effects like soft gluon emission.
\item[2.] ``$X_{\rm extra Size}$'' is a parameter which allows the user to
  increase homogeneously the overall number of $x$ values on the
  interval $[10^{-7}:1-10^{-7}]$ {\it after} the first program
  DGLAP\_MSSM (which is, once again, the most time consuming part of
  the complete code). But it is quite {\bf useless}, the initial value
  of $X_{\rm Size}$ being large enough for all following
  programs\footnote{In fact, this is not exactly true, because the
    implementation of 2-body decays sometimes requires a local
    increase of the precision, and thus a local $x$ array. But this is
    fully implemented in the programs themselves, and is totally
    hidden from the user.}. By default, $X_{\rm extra Size}$ is simply
  taken to be equal to $X_{\rm Size}$. Of course, it has to be greater
  than (or at least equal to !) $X\_{\rm Size}$.
\item[3)] ``Part\_init'' and ``Part\_fin'' describe the initial and
  final id's of an {\it interval of initial particles} for which the
  FFs have to be computed. Note that there are 30 initial particles in
  the MSSM (see Appendix A for the description of these particles and
  their id's), and {\it all} the $30 \times 30 = 900$ FFs from any
  particle to any other will be needed for the computation of the
  whole cascade. Thus the default values are respectively Part\_init $ =
  1$ and Part\_fin $= 30$, which means that the program will treat
  successively all the 30 possible initial particles. Nevertheless,
  the treatment of an initial particle being fully independent of the
  others, any of the 3 programs DGLAP\_MSSM, Susy1TeV and Less1GeV can
  be cut into pieces to be run independently on different computers;
  for example, you can let a first computer run the chosen program for
  particles 1 to 15, and another computer run the {\it same} program
  for particles 16 to 30. These two parameters render this task easy
  and allow to save a lot of time.\newline {\bf Caution}: Note that each of
  these three programs has to be run over the {\it whole range of
    particles} before running the following one !
\item[4.] $X_{\rm min}$ gives the lowest value of the final x interval. As
  stated above, in the lowest x region ($[X_{\rm min}:10^{-7}]$), you can
  choose two different extrapolations of the FFs : either
  extrapolating the power law obtained from the LO DGLAP equations, or
  using the MLLA approximation for taking color coherence effects into
  account. This parameter, taken by default to be $X_{min} =
  10^{-13}$, is only used in the very last part of the computation of
  the cascade : Less1GeV.\newline
{\bf Caution}: of course, $X_{\rm min}$ has to be $\ge 10^{-7}$.

\end{itemize}

\subsection{DGLAP\_MSSM}
\label{subsec:DGLAP_MSSM}

\setcounter{footnote}{1}

This program treats completely the perturbative cascade above the
$M_{\rm SUSY}$ scale. Starting from input FFs at $M_{\rm SUSY}$ for each type
of primary particle $P$ ($D_P^P(x,M_{\rm SUSY}) = \delta(1-x)$ and $\forall
j \neq P$, $D_P^j(x,M_{\rm SUSY}) = 0)$, it gives the FFs of the 30
interaction eigenstates at scale $Q = M_X$: $D_P^j(x,M_X)$.

By giving the parameters of the ``Input.dat'' file, the user can
choose one of the 4 available theories, namely 1 : MSSM, 2 : SM, 3 :
SUSY-QCD, 4 : QCD alone. I point out that this complicated program is
certainly not the best one for treating a case as simple as QCD DGLAP
equations (or even SUSY-QCD), being unfortunately quite time
consuming. This program requires no external input (except
Input.dat, of course), and only needs as ``physical inputs'' the
values of $\beta$ and $M_X$, described above. The technical parameters
$X_{\rm Size}$, ``Part\_init'' and ``Part\_fin'' are used, too. As I
already mentioned before, I strongly suggest when possible to run the
program on different computers at the same time, using different
intervals of initial particles, for saving time\footnote{Again,
  the running time will depend on the computer you are using. Yet, to
  give an idea, you should foresee around one hour of running time per
  initial particle in the MSSM framework for $M_X \sim 10^{16}$ GeV on
  a modern computer (1 GHz or more, 256 Mo of RAM).}.

Finally, the user should specify the corresponding output directory,
where the output files will be stored.

Using the structure of the DGLAP evolution equations and
$\delta$-functions input FFs at $M_{\rm SUSY}$ (practically implemented as
sharp gaussians), this program will compute the full set of FFs from
one particle to another between $M_{\rm SUSY}$ and $M_X$. For this
purpose, I use a Runge-Kutta method with a {\it constant logarithmic}
step in virtuality for solving the system of DGLAP
equations\footnote{Unfortunately, for practical reasons, it was not
possible to choose a floating step.}. There must be an {\bf entire
number} of these steps between $M_{SUSY}$ and (any value of)
$M_X$. That's why it is only possible to use powers of 10 for these
scales. Nevertheless, as I said before, this allows already a good
accuracy.

Here we can see the interest of the variable
``Nb\_output\_virtualities'' and the corresponding array of virtuality
values ``Output\_virtualities\_DGLAP\_MSSM'' : thanks to the fact that
this program is computing the FFs from $M_{SUSY}$ to $M_X$ through a
given number of Runge-Kutta steps, all intermediate virtualities used
by the Runge-Kutta program are available as possible outputs ; it
allows to get the FFs at intermediate virtualities, which are equivalent
to lower $X$ masses $M_X$. As stated above, the step used for
Runge-Kutta is a constant logarithmic step, exactly one order of
magnitude each. So practically, the user who wants to study a GUT X
particle with mass $M_X \sim 10^{25}$ eV can get the results for any
other (power of 10) $X$ mass ($10^{21}$, $10^{22}$, $10^{23}$ eV,...)
between $M_{SUSY}$ and $M_X$. The two variables cited above allow to
put these partial results in output files that will be usable later
on. Note again that only {\it this} program will use the array of values for
$M_X$. The following ones will simply use one of these values, the one
given in the parameter $M_X$ itself.

The output is presented in $30 \times 30 = 900$ files giving the FFs
of any particle into any other with $X_{\rm Size}$ values of x in the
first column and the corresponding values of the FF in the second one.
These files are called generically ``fragment\_(M\_X)eV\_p1.p2'' for
the FF of particle p1 into particle p2, where (M\_X) contains the mass
of the $X$ particle at which the FF was computed\footnote{Note that,
  according to the form of DGLAP equations, the iterator
  part$\in[$Part\_init,Part\_fin$]$ of the program DGLAP\_MSSM (and
  evidently its ``brother'' DGLAP\_QCD) runs over the ``final''
  particles p2. On the contrary, the equivalent iterators in Susy1TeV
  and Less1GeV run over particles p1. That's why it is {\it essential}
  to run the different programs successively, after the complete end
  of the preceding one !}.

Finally, it is worth noting that the output of this program
only depends on very few parameters : the $M_{\rm SUSY}$ scale at which
the perturbative cascade is ending, and the SUSY parameter $\tan
\beta$. But as stated in \cite{bd2}, these two parameters
have very little influence on the final results\footnote{Indeed, the
evolution being only logarithmic in virtuality, and running over many
orders of magnitude until $M_X$, the exact value of $M_{SUSY}$ (say,
between 200 GeV and 1 TeV) doesn't really matter. Similarly, the $\tan
\beta$ parameter only affects the Yukawa interactions, which are
almost negligible, except in some rare cases for the third generation
of quarks and leptons.}. Thus I strongly recommend to let run this
program only {\it once}, for all $X$ masses you want to study, and to
carefully keep these partial results for later use, for studying the
influence of other parameters appearing in the following programs.

\subsection{How to use Isasusy.x}
\label{subsec:Isasusy}

The ISASUSY program, written in fortran 77, is a subset of the whole
code called ISAJET. I refer to the user manual of
ISASUSY\cite{Isasusy} for information about how to use this program.
But, as mentioned above, I fully implemented a personalised version of
this code, which is available through the master program (option 7).
For a given set of SUSY parameters specified by the user in SUSY.dat,
it computes the complete SUSY spectrum (masses and mixing angles of
all the sparticles, stored in the ``Mixing.dat'' file), and the
allowed decay modes with the corresponding branching ratios (stored in
``Decay.dat''). Both files will be stored in the Isasusy directory of
SHdecay, and their names have to be given in the ``SUSY.dat'' input
file. 

Of course, you can get these files from any other available code
providing the same information as ISASUSY, as long as you adapt the
output format of this code in order to get the one required by SHdecay
(see the model files provided in the Isasusy directory for information
about the required format).  The furnished version of Isasusy is the
one included in Isajet 7.51. If you want to use an updated version of
Isasusy, you probably just need to replace the two files called
``aldata.f'' and ``libisajet.a'' in the Isasusy repertory of SHdecay
(but hopefully {\it not} ssrun.f and the Makefile, which I have
adapted).Yet, I obviously cannot ensure that this operation will
work...

\subsection{Susy1TeV}
\label{subsec: Susy1TeV}

This program takes the results of DGLAP\_MSSM given at the ({\it
unique}) $M_X$ value specified in Input.dat and deals with the
breaking of SUSY and $SU(2) \otimes U(1)$, the supersymmetric decay
cascade and the decays of the top quarks, the Higgs, $W$ and $Z$ bosons.
The muons and taus existing at this step are decayed too. I
considered only 2- and 3-body decays for which I computed the
relevant phase space\footnote{We compute these $N$-body decays from phase
space, including all mass effects, but I didn't include the matrix
elements.}, using the branching ratios and the mass spectrum given by
ISASUSY. For any detail on these procedures, I refer to \cite{bd2}.

The input directory has to be the one where the outputs of DGLAP\_MSSM
have been stored (the user doesn't have to specify it). On the other
hand, the output directory can be different, in order to {\it
  distinguish between different parameters}. For example, the mass of
the $X$ particle, which is especially specified in the names of the
output files of DGLAP\_MSSM (as being the final virtuality of the
perturbative cascade) is {\it no more specified} in the outputs of
Susy1TeV ; thus it can be usefull to define different output
directories for different $X$ masses. Moreover, the large number of
output files is easier to handle when stored in different
directories.

The program will need the two output files given by Isasusy.x (or any other
program, see above); the two files have to be located in the
directory ``Isasusy'', and their names have to be given in the two
corresponding parameters of SUSY.dat : ``Decay'' and ``Mixing''. Here
the parameter ``Region'' also becomes useful.  If necessary, the
extension of the $x$ array to ``$X_{\rm extra Size}$'' values instead of
``$X_{\rm Size}$'' will occur in this program, too\footnote{Though, as
  I already mentioned, this option is not of very big use.}.

The output files contain the FFs of the 30 initial particles
(interaction eigenstates) into the remaining SM mass eigenstates after
the decays, namely the quarks u, d, s, c, b and gluons, the electrons,
neutrinos and the LSP. All of them will have a suffix ``\_1TeV'' to
distinguish them from the outputs of other programs and a
``region''-suffix, e.g. labelling the set of SUSY parameters you used
during the run.

\subsection{DGLAP\_QCD}
\label{subsec:DGLAP_QCD}

This program is a simplified copy of DGLAP\_MSSM. It computes the pure
QCD perturbative partonic cascade for quarks\footnote{Only 5 quarks
  are considered here, namely $u,d,s,c,b$, the top quarks having been
  decayed at scale $M_{\rm SUSY}$.} and gluons (so only 6 particles)
for a virtuality decreasing from $M_{\rm SUSY}$ to $Q_{\rm had} =
{\rm max}(m_{\rm quark},1$ GeV). This program is {\it not using} any previous
result from other ones, and only depends on $M_{\rm SUSY}$ and $Q_{\rm had}$,
which are not very sensitive parameters, as stated above. I thus
recommend to define their values once and for all (say, keep the default
values $M_{\rm SUSY} = 1 $ TeV and $Q_{\rm had} = 1$ GeV), and to run
DGLAP\_QCD {\it only once}. This possiblity of sparing running time is
the reason why the necessary convolution between the results of this
program and the FFs given by the previous one (Susy1TeV) was
implemented in Less1GeV, in order to keep DGLAP\_QCD fully
independent.

The $6 \times 6 = 36$ output files, called generically ``fragment\_p1.p2'' -
where p1 and p2 are initial and final partons \{u,d,s,c,b,g\} - will
be stored in the corresponding directory given in Input.dat. I
recommend to use a dedicated directory, for the reason stated above :
these results are almost parameter independent and can be used for
different runs of Susy1TeV and Less1GeV.

\subsection{Fragment\_maker}
\label{subsec:Fragment}

This program is certainly the weakest part of our treatment, because
of the lack of knowledge concerning the non-perturbative FFs at very
low $x$. I used the results of \cite{Poetter} for this purpose, which
are based on LEP data. Unfortunately they are only valid for
$x \geq 0.1$.  The reason is that at LEP energies, it is
necessary to consider mass effects at small $x$, which can be described
by the so-called ``MLLA plateau''. Such effects can be taken into
account during the computation of the hadronization itself, in
Less1GeV. In Fragment\_maker,  just keep the FFs given in
\cite{Poetter} up to $x = 0.1$ and extrapolate them at small $x$, by
requiring continuity and the overall conservation of energy. 

We finally obtain a set of input functions for light quarks (including
the $b$) and for gluons which conserve energy and agree with known
data.

This program is fully independent of the others, and is just used to
``prepare'' the non-perturbative input FFs at low energy needed in
Less1GeV. It doesn't depend on any parameter, and can be run once and for
all. Once again, I recommend to use a dedicated directory for storing the
output Files of this program ; the default value is a directory called
``Fragment''.

\subsection{Less1GeV}
\label{subsec:Less1GeV}

This program first computes the convolution between the results of
Susy1TeV (describing the evolution of the FFs between $M_X$ and
$M_{\rm SUSY}$ after SUSY, top, $W$, $Z$ and Higgs decays) and the ones of
DGLAP\_QCD (describing the further evolution of the partonic part of
these FFs between $M_{\rm SUSY}$ and $Q_{\rm had}$). It further deals with the
hadronization of quarks and gluons, using external input FFs (The
results of the Fragment\_maker program described above) which have to
be convoluted with the previous results. It finally deals with the
decays of the last unstable particles. The 2- and 3-body decays are
treated exactly as in Susy1TeV.

The results are once more given in terms of FFs of any initial
(interaction eigenstate) particle (between the 30 available in the
MSSM, see Appendix A) into the final (physical) stable ones, namely
the protons, electrons, photons, three species of neutrinos, and LSPs.
To simplify the storing and further use of these (final) results, I
grouped all the results corresponding to one initial particle in one
file generically called ``fragment\_p1.all\_Region'', where p1 is the
initial particle and Region the suffix labelling the set of SUSY
parameters. Each file contains seven FFs : the first column gives the
values of $x$ (from $1-10^{-7}$ to $X_{\rm min}$, in decreasing order),
and the next columns give successively the FFs of p1 into protons,
photons, LSPs, electrons, $\nu_e$, $\nu_\mu$, $\nu_\tau$.

\subsection{Xdecay}
\label{subsec:Xdecay}

This last program allows to study a special decay mode of the X
particle, by computing a last convolution between the results obtained
in Less1GeV and the phase space of the given decay mode. The number of
decay products and their nature (through the associated id, see
Appendix A) have to be specified in Input.dat. 

If a decay mode for the $X$ particle has been specified in the two
parameters ``N-body\_X\_decay'' and ``X\_decay\_mode'' (respectively
the number $N$ of products and the id's associated to each product - see
Appendix A), a last convolution with the $N$-body decay energy spectrum
will be computed and the results will be directly given in terms of
the FFs of the $X$ particle into the stable final ones. The $N$-body
energy spectrum I used is the one given in \cite{Sarkar:2001}. If
$\rho_N(z)$ is the probability density of obtaining a decay product of
energy $E$ carrying the energy fraction $x = 2E/M_X$ of the decaying
particle, we have :

\beqa
\label{e9}
&& \rho_2(x) = \delta(1-x)\,,\nonumber\\
&& \rho_N(x) = (N-1)(N-2)x(1-x)^{N-3}\,, N\geqslant 3\,.\nonumber\\
\eeqa

This program has been separated from Less1GeV to allow the user to
obtain very quickly any decay mode he wants to study. The final result
is stored in the same directory as the results of Less1GeV. It is
generically called ``frag\_X\_a\_b\_c.all\_G'' and has the same format
as the one described above for the results of Less1GeV.

\section{Conclusion}

This article describes in some detail how to use the code SHdecay,
which has been designed for computing the most general decay spectra
of any super-heavy particle in the framework of the MSSM. I hope that
it will be of some use for other researchers. The code is available on
the web site of our group, under the address
:''www1.physik.tu-muenchen.de/~barbot/'', and I will be pleased to
answer any question you have about it. Of course, any remark or
suggestion is welcome, too.

\section*{Appendix A: Description of the compound particles used in SHdecay}

\setcounter{footnote}{1}

Here I describe the 30 interaction eigenstates (or ``compound
particles'') of the MSSM which have been used as possible decay
products for the $X$ particle. As the decay is occuring well above the
breaking scales of SUSY and $SU(2) \otimes U(1)$, one has to allow a
decay into supersymmetric particles as well as SM particles, and to
distinguish between the helicities (Left or Right) of the Dirac
fermions; yet, well above the breaking scales of SUSY and $SU(2)
\otimes U(1)$, it is assumed that one doesn't need to distinguish
between the components of a given SU(2) multiplet\footnote{This is
  certainly true if $X$ is an $SU(2)$ doublet.}, in particular between
the ``up'' and ``down'' components of the SU(2) doublets. Moreover, up
to the Yukawa couplings which become relevant only for the third
generation of fermions, no difference is made between the generations,
all particles being massless above the $SU(2) \otimes U(1)$ breaking
scale. If we consider a perfect CP symmetry, one doesn't need to
distinguish between particles and antiparticles, either. In summary,
for example, the fields ($u_L$, $d_L$), ($c_L$, $s_L$), ($\bar{u}_L$,
$\bar{d}_L$), and ($\bar{c}_L$, $\bar{s}_L$) all obey exactly the same
DGLAP evolution equation and thus can be considered as a {\it single}
``particle'' which is taken to be an {\it average} over all these
fields. This ``coumpound particle'' is called $q_L$ in our
nomenclature and has id 1. I give in table~\ref{fermions} all
fermionic compound particles I used, together with the associated
superparticles, and their respective id's.

The same occurs for bosons and bosinos, where we only have to consider
the unbroken fields $B$, $W$, $g$ (for gluons), the two SU(2) Higgs doublets
of the MSSM $H_1$ (coupled to leptons and down-type quarks of the
third generation) and $H_2$ (coupled to the up-type quarks of the
third generation), and their superpartners. The well known particles
and antiparticles at lower energies are mixtures of the components
of these interaction eigenstates. I give the corresponding id's in
table~\ref{bosons}.


\section*{Test Run Input and Output} 
\label{sec:io}

I give here a test run for a very simple case that can be computed in
a few hours: the ``Input\_ex.dat'' file and the final ``Output\_ex.dat''
file given by SHdecay for a $X$ particle with $M\_X = 10^{10}$ GeV in
the pure QCD case (Theory = 4).

\medskip
\begin{center}
{\bf {\large \underline{Input\_ex.dat:}}}
\end{center}
\medskip

{\bf

Theory: 4

MLLA: 1

Nb\_output\_virtualities: 1

Output\_virtualities\_DGLAP\_MSSM(GeV): 1.e10

M\_x(GeV): 1.e10

N-body\_X\_decay: 2

X\_decay\_mode: 1 1

M\_Susy: 1.e3

Q\_had: 1.

X\_Size: 101

X\_extra\_Size: 101

Part\_init: 1

Part\_fin: 30

X\_min: 1.e-13

Region: G

Output\_file: ./QCD/Output\_ex.dat

Output\_directory(DGLAP\_MSSM.c): QCD

Output\_directory(Susy1TeV.c):    QCD

Output\_directory(DGLAP\_QCD.c):    QCD/DGLAP\_QCD

Output\_directory(Fragment\_maker.c): Fragment

Output\_directory(Less1GeV.c):    QCD}

\bigskip

\begin{center}
{\bf {\large \underline{Output\_ex.dat:}}}
\end{center}

\medskip

{\bf
********************** Input Parameters **********************\\

\noindent
THEORY = 4 (1 for MSSM, 2 for SM, 3 for Susy-QCD, 4 for QCD alone)\\
MLLA = 1 (1 for MLLA approximation, 0 for power laws at small x)\\
Virtualities at which DGLAP\_MSSM is storing results in output files :
  Mx[0] = 1e+10 GeV\\
Mass of the X particle (it must be one of the virtualities at which results have been stored !):  Mx = 1e+10 GeV\\
Number of decay products of the X particle : 2\\
Decay mode of the X particle :1  1  \\
Susy scale : 1000 GeV\\
Hadronization scale : 1 GeV\\
Beta = 1.29985\\

\noindent
X\_Size (in DGLAP\_MSSM): 101\\
X\_extra\_Size (everywhere else) : 101\\
Xmin : 1e-13\\
Computation from part 1 to part 30\\
Region : G (For example, G for Gaugino-like LSP, H for Higgsino-like LSP)\\

\noindent
DGLAP\_MSSM output directory = QCD\\
Susy1TeV output directory = QCD\\
DGLAP\_QCD output directory = QCD/DGLAP\_QCD\\
Fragment\_maker output directory = Fragment\\
Final output directory = QCD\\

\medskip

**********************  Susy1TeV  ***************************\\

                    Initial Particle 1 = uL\\\indent
                  -----------------------------\\

\noindent
                  uL $->$ quarks + gluons = 0.998024\\
                  uL $->$ gamma = 0\\
                  uL $->$ LSP = 0\\
                  uL $->$ e = 0.000174832\\
                  uL $->$ Nu\_e = 0.000174743\\
                  uL $->$ nu\_mu = 0.000174463\\
                  uL $->$ nu\_tau = 0.00016671\\
                  uL $->$ Total = 0.998715\\

\clearpage

[...]

***********************  DGLAP\_QCD  ************************ \\

[...]\\

                  PARTON = 5\\\indent
               ------------------\\

\noindent
At final t: \\

\noindent
Partial energy from 0 into 5 = 0.0502139\\
Partial energy from 1 into 5 = 0.00935672\\
Partial energy from 2 into 5 = 0.00935672\\
Partial energy from 3 into 5 = 0.00935672\\
Partial energy from 4 into 5 = 0.00935672\\
Partial energy from 5 into 5 = 0.69335\\

\noindent
Integrated energy from 0 = 0.997878\\
Integrated energy from 1 = 0.998345\\
Integrated energy from 2 = 0.998345\\
Integrated energy from 3 = 0.998345\\
Integrated energy from 4 = 0.998339\\
Integrated energy from 5 = 0.998346\\

\medskip

**********************  Fragment\_maker  **********************\\

Total energy fractions contained in the extrapolated FFs (including the hadronic Peterson contribution for heavy quarks c and b) :\\
(should be very close to 1. !)\\

\noindent
   Energy fraction [u] = 0.997544\\
   Energy fraction [d] = 0.997544\\
   Energy fraction [s] = 0.996595\\
   Energy fraction [c] = 1.07197\\
   Energy fraction [b] = 1.032\\
   Energy fraction [g] = 1.00657\\

\clearpage

**********************  Less1GeV  ************************** \\

                    Initial Particle 1 = uL\\\indent
                  -----------------------------\\

\noindent
                  uL $->$ protons = 0.112688\\
                  uL $->$ gamma = 0.253301\\
                  uL $->$ LSP = 0\\
                  uL $->$ e = 0.162863\\
                  uL $->$ Nu\_e = 0.160729\\
                  uL $->$ nu\_mu = 0.3005\\
                  uL $->$ nu\_tau = 0.000698547\\
                  uL $->$ Total = 0.990781\\

[...]\\

\medskip

**********************  X\_decay  *************************** \\

Total fraction of energy carried by decay product 1 = 0.990781\\

(This result should be close to 1., in order to respect the energy conservation over the whole cascade)\\

Total fraction of energy carried by decay product 0 = 0.990781\\

(This result should be close to 1., in order to respect the energy conservation over the whole cascade)\\

Energy fraction carried by the X particle = 1.98156\\

(This result should be almost equal to 2, because the mass of the X particle is twice the energy fraction provided to any of its decay products. In this final result, the total energy carried by all FFs is naturally equal to the mass of the initial particle !)\\

\bigskip
*************************************************************
}
\bigskip

\begin{figure}
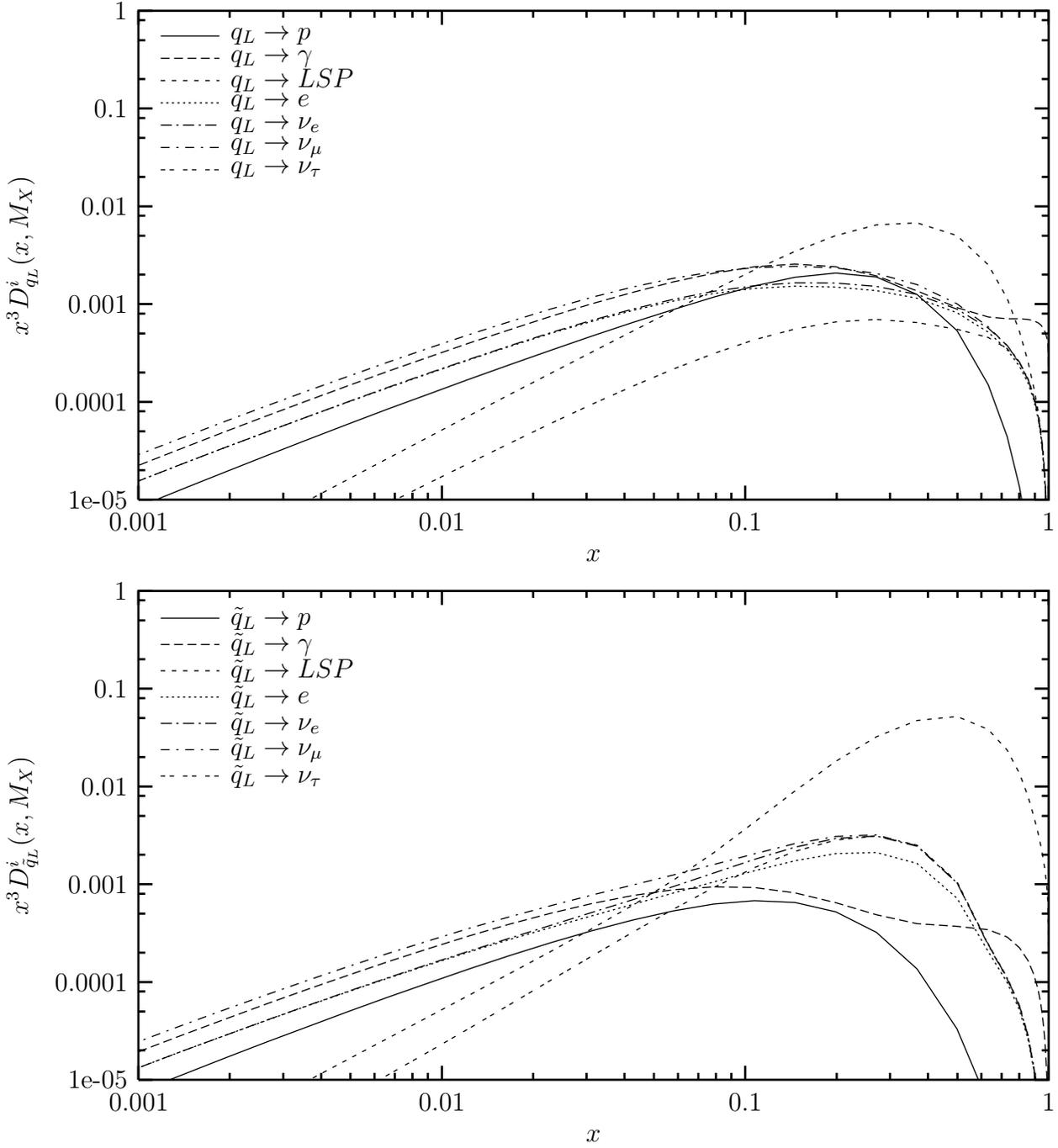

\label{fig:FF}
\input{Low_G_uL.tex}
\input{Low_G_uL_.tex}
\caption{Fragmentation functions of a first or second generation
  $SU(2)$ doublet quark (top) and a squark (bottom) into stable
  particles.}
\end{figure}

In order to illustrate the type of results that one can get in the
most general case (MSSM framework, Theory = 1), I give in
fig.~\ref{fig:FF} an example of output fragmentation functions
obtained with SHdecay for initial $u_L$ quark (id.~1 in Appendix A)
and $\tilde{u}_L$ squark (id.~7 in Appendix A), for one set of SUSY
parameters, with low $\tan\beta$ and gaugino--like LSP. I used the
default values of the furnished files Input.dat and SUSY.dat (i.e. a
ratio of Higgs vevs $\tan \beta = 3.6$, a gluino and scalar mass scale
$M_{\rm SUSY} \sim 1000$ GeV, a supersymmetric Higgs mass parameter
$\mu = 500$ GeV, a CP--odd Higgs boson mass $m_A = 1000$ GeV and
trilinear soft breaking parameter $A_t = 1$ TeV). As usual, I plot
$x^3 \cdot D_I^P (x,M_X)$. I take $M_X = 10^{16}$ GeV, as appropriate
for a GUT interpretation of the $X$ particle.

\clearpage


\begin{center}
\begin{table}[ht!]
\begin{center}
\begin{tabular}{|c|c|}\hline
compound particle &\rule[-3mm]{0mm}{8mm}id\\\hline
$q_L = \frac{1}{4}\left[\left(\frac{u_L}{d_L}\right) + \left(\frac{c_L}{s_L}\right) +
    \left(\frac{\bar{u}_L}{\bar{d}_L}\right) +
    \left(\frac{\bar{s}_L}{\bar{c}_L}\right)\right]$&1\\
$q_R = \frac{1}{4}\left[u_R + c_R + \bar{u}_R + \bar{c}_R \right]$&2\\
$d _R = \frac{1}{4}\left[d_R + s_R + \bar{d}_R + \bar{s}_R \right]$&3\\
$t_L = \frac{1}{2}\left[\left(\frac{t_L}{b_L}\right) +
  \left(\frac{\bar{t}_L}{\bar{b}_L}\right) \right]$&4\\
$t_R = \frac{1}{2}(t_R + \bar{t}_R)$&5\\
$b_R = \frac{1}{2}(b_R + \bar{b}_R)$&6\\
&\\
$\tilde{q}_L = \frac{1}{4}\left[\left(\frac{\tilde{u}_L}{\tilde{d}_L}\right) + \left(\frac{\tilde{c}_L}{\tilde{s}_L}\right) +
    \left(\frac{\tilde{\bar{u}}_L}{\tilde{\bar{d}}_L}\right) +
    \left(\frac{\tilde{\bar{s}}_L}{\tilde{\bar{c}}_L}\right)\right]$&7\\
$u_R = \frac{1}{4}\left[\tilde{u}_R + \tilde{c}_R + \tilde{\bar{u}}_R + \tilde{\bar{c}}_R \right]$&8\\
$d _R = \frac{1}{4}\left[\tilde{d}_R + \tilde{s}_R + \tilde{\bar{d}}_R + \tilde{\bar{s}}_R \right]$&9\\
$t_L = \frac{1}{2}\left[\left(\frac{\tilde{t}_L}{\tilde{b}_L}\right)+
  \left(\frac{\tilde{\bar{t}}_L}{\tilde{\bar{b}}_L}\right) \right]$&10\\
$t_R = \frac{1}{2}(\tilde{t}_R + \tilde{\bar{t}}_R)$&11\\
$b_R = \frac{1}{2}(\tilde{b}_R + \tilde{\bar{b}}_R)$&12\\
&\\
$l_L = \frac{1}{4}\left[\left(\frac{e_L}{\nu_e}\right) + \left(\frac{\mu_L}{\nu_\mu}\right) +
    \left(\frac{\bar{e}_L}{\bar{\nu}_e}\right) +
    \left(\frac{\bar{\mu}_L}{\bar{\nu}_\mu}\right)\right]$&13\\
$l_R = \frac{1}{4}\left[e_R + \mu_R + \bar{e}_R + \bar{\mu}_R
\right]$&14\\
$\tau_L = \frac{1}{2}\left[\left(\frac{\tau_L}{\nu_\tau}\right) +
  \left(\frac{\bar{\tau}_L}{\bar{\nu}_\tau}\right) \right]$&15\\
$\tau_R = \frac{1}{2}(\tau_R + \bar{\tau}_R)$&16\\
&\\
$\tilde{l}_L = \frac{1}{4}\left[\left(\frac{\tilde{e}_L}{\tilde{\nu}_e}\right) + \left(\frac{\tilde{\mu}_L}{\tilde{\nu}_\mu}\right) +
    \left(\frac{\tilde{\bar{e}}_L}{\tilde{\bar{\nu}}_e}\right) +
    \left(\frac{\tilde{\bar{\mu}}_L}{\tilde{\bar{\nu}}_\mu}\right)\right]$&17\\
$l_R = \frac{1}{4}\left[\tilde{e}_R + \tilde{\mu}_R +
  \tilde{\bar{e}}_R + \tilde{\bar{\mu}}_R \right]$&18\\
$\tau_L = \frac{1}{2}\left[\left(\frac{\tilde{\tau}_L}{\tilde{\nu}_\tau}\right) +
  \left(\frac{\tilde{\bar{\tau}}_L}{\tilde{\bar{\nu}}_\tau}\right) \right]$&19\\
$\tau_R = \frac{1}{2}(\tilde{\tau}_R + \tilde{\bar{\tau}}_R)$&20\\\hline
\end{tabular}
\caption{Definition and id's of the compound SM fermions and their superpartners in SHdecay.}
\label{fermions}
\end{center}
\end{table}
\end{center}

\begin{center}
\begin{table}[ht!]
\begin{center}
\begin{tabular}{|c|c|}\hline
compound particle &\rule[-3mm]{0mm}{8mm}id\\\hline
$W = \frac{1}{3}(W_1 + W_2 + W_3)$&21\\
$B$&22\\
$g$&23\\
$H_1 = \frac{1}{2}(H_1^1+H_1^2)$&24\\
$H_2 = \frac{1}{2}(H_2^1+H_2^2)$&25\\
&\\
$\tilde{W} = \frac{1}{3}(\tilde{W}_1 + \tilde{W}_2 + \tilde{W}_3)$&26\\
$\tilde{B}$&27\\
$\tilde{g}$&28\\
$\tilde{H}_1 = \frac{1}{2}(\tilde{H}_1^1+\tilde{H}_1^2)$&29\\
$\tilde{H}_2 = \frac{1}{2}(\tilde{H}_2^1+\tilde{H}_2^2)$&30\\\hline
\end{tabular}
\caption{Definition and id's of the compound bosonic SM particles and their superpartners in SHdecay.}
\label{bosons}
\end{center}
\end{table}
\end{center}


\vspace*{5mm}
\begin{center} 
\begin{figure}[h!]
\begin{picture}(-400,300)(0,-190)
\label{X_decay}
\SetOffset(500,0)
\SetPFont{Helvetica}{24}
\PText(-405,4)(0)[]{X}
\DashLine(-410,0)(-440,0){3} \Text(-425,7)[]{$\tilde{q}_L$} 
\DashLine(-400,0)(-370,0){3} \Text(-385,7)[]{$\tilde{q}_L$} 
\Line(-370,1)(-340,26)
\Line(-370,0)(-340,25) \Text(-360,25)[]{$\tilde{g}$} 
\Line(-340,26)(-310,46)
\Line(-340,25)(-310,45) \Text(-330,45)[]{$\tilde{g}$}
\Gluon(-340,25)(-310,10){-3}{4}\Text(-320,25)[]{$g$} 
\DashLine(-310,45)(-280,55){3}  \Text(-295,60)[]{$\tilde{q}_L$} 
\GCirc(-280,55){3}{0}
\Line(-310,45)(-280,35) \Text(-295,30)[]{$q_L$} 
\DashLine(-280,-120)(-280,110){4} \Text(-280,-130)[c]{1 TeV}
\Text(-280,-145)[c]{(SUSY}
\Text(-280,-160)[c]{+ $SU(2)\otimes U(1)$}
\Text(-280,-175)[c] {breaking)}

\Line(-280,55)(-250,80) \Text(-265,80)[]{$q$} 
\Line(-250,80)(-190,70) \Text(-220,68)[]{$q$} 
\Gluon(-250,80)(-220,90){3}{4} \Text(-235,97)[]{$g$} 
\Line(-220,90)(-190,100) \Text(-205,103)[]{$q$} 
\Line(-220,90)(-190,80) \Text(-205,80)[]{$q$} 

\Line(-280,56)(-250,46)
\Line(-280,55)(-250,45)
\Text(-265,42)[]{$\tilde{\chi}_2^0$}
\GCirc(-250,45){3}{0}
\Line(-250,45)(-220,55) \Text(-235,57)[]{$q$}
\Line(-250,46)(-170,36)
\Line(-250,45)(-170,35) \Text(-166,35)[l]{$\tilde{\chi}_1^0$}
\Line(-250,45)(-220,25) \Text(-235,27)[]{$q$} 
\Line(-220,55)(-190,65) 
\Gluon(-220,55)(-190,45){-3}{4} 
\DashLine(-190,-120)(-190,110){4} \Text(-190,-130)[c]{1 GeV}
\Text(-190,-145)[c]{(hadronization)}

\Line(-370,0)(-340,-25) \Text(-360,-18)[]{$q_L$} 
\Photon(-340,-25)(-310,-5){3}{5} \Text(-330,-5)[]{$B$} 
\DashLine(-310,-5)(-280,5){3} \Text(-295,8)[]{$\tilde{q}_R$}
\GCirc(-280,5){3}{0}
\DashLine(-310,-5)(-280,-15){3} \Text(-295,-18)[]{$\tilde{q}_R$}
\GCirc(-280,-15){3}{0}
\Line(-340,-25)(-310,-45) \Text(-330,-40)[]{$q$} 
\Photon(-310,-45)(-280,-25){3}{5} \Text(-308,-31)[]{$W$} 
\GCirc(-280,-25){3}{0}
\Line(-280,-25)(-190,-5) \Text(-240,-8)[]{$\tau$} 
\GCirc(-190,-5){3}{0}
\Line(-190,-5)(-160,5) \Text(-175,9)[]{$a_1^{-}$}
\GCirc(-160,5){3}{0}
\Line(-160,5)(-130,15) \Text(-145,17)[]{$\rho^{-}$}
\GCirc(-130,15){3}{0}
\Line(-130,15)(-100,15) \Text(-115,11)[]{$\pi^{-}$}
\GCirc(-100,15){3}{0}
\Line(-100,15)(-70,15) \Text(-66,15)[l]{$\nu_\mu$} 
\Line(-100,15)(-70,5) \Text(-78,2)[]{$\mu^{-}$} 
\GCirc(-70,5){3}{0}
\Line(-70,5)(-50,5) \Text(-46,5)[l]{$\nu_\mu$} 
\Line(-70,5)(-50,-5) \Text(-46,-7)[l]{$\nu_e$} 
\Line(-70,5)(-50,-15) \Text(-46,-16)[l]{$e^{-}$} 

\Line(-130,15)(-100,35) \Text(-115,35)[]{$\pi^0$} 
\GCirc(-100,35){3}{0}
\Photon(-100,35)(-70,35){3}{5} \Text(-66,35)[]{$\gamma$} 
\Photon(-100,35)(-70,55){3}{5} \Text(-66,57)[]{$\gamma$} 

\Line(-160,5)(-130,-5) \Text(-145,-8)[]{$\pi^0$}
\GCirc(-130,-5){3}{0}
\Photon(-130,-5)(-100,-25){3}{5} \Text(-96,-27)[]{$\gamma$} 
\Photon(-130,-5)(-100,-5){3}{5} \Text(-96,-3)[]{$\gamma$} 
\Line(-190,-5)(-160,-15) \Text(-175,-16)[]{$\nu_\tau$} 
\Line(-280,-25)(-100,-45) \Text(-96,-45)[l]{$\nu_\tau$} 

\Line(-310,-45)(-280,-65) \Text(-300,-62)[]{$q$} 
\Line(-280,-65)(-250,-55) \Text(-265,-53)[]{$q$} 
\Gluon(-280,-65)(-250,-75){-3}{4} \Text(-265,-80)[]{$g$}
\Gluon(-250,-75)(-220,-95){-3}{4} \Text(-240,-95)[]{$g$} 
\Line(-220,-95)(-190,-105) \Text(-205,-105)[]{$q$} 
\Line(-220,-95)(-190,-85) \Text(-205,-85)[]{$q$}

\Gluon(-250,-75)(-220,-65){3}{4} \Text(-240,-62)[]{$g$} 
\Line(-220,-65)(-190,-55)  \Text(-205,-55)[]{$q$}
\Line(-220,-65)(-190,-75)  \Text(-205,-75)[]{$q$}

\GOval(-190,65)(20,7)(0){0} 
\GCirc(-170,65){3}{0}
\Text(-175,65)[r]{$n$}
\Line(-170,65)(-130,80) \Text(-126,82)[l]{$p$}
\Line(-170,65)(-130,65) \Text(-126,65)[l]{$e^{-}$}
\Line(-170,65)(-130,50) \Text(-126,48)[l]{$\nu_e$} 

\GOval(-190,-80)(15,7)(0){0} 
\GCirc(-168,-83){3}{0}
\Text(-170,-80)[r]{$\pi^0$}
\Photon(-168,-83)(-130,-65){3}{5} \Text(-126,-63)[]{$\gamma$} 
\Photon(-168,-83)(-130,-95){3}{5} \Text(-126,-98)[]{$\gamma$} 

\end{picture} 
\caption{Schematic MSSM cascade for an initial squark with a
virtuality $Q \simeq M_X$. The full circles indicate decays of massive
particles, in distinction to fragmentation vertices. The two vertical
dashed lines separate different epochs of the evolution of the
cascade: at virtuality $Q > M_{\rm SUSY}$, all MSSM particles can be
produced in fragmentation processes. Particles with mass of order
$M_{\rm SUSY}$ decay at the first vertical line. For $M_{\rm SUSY} > Q
> Q_{\rm had}$ light QCD degrees of freedom still contribute to the
perturbative evolution of the cascade. At the second vertical line,
all partons hadronize, and unstable hadrons and leptons decay. See the
text for further details.}
\end{figure}
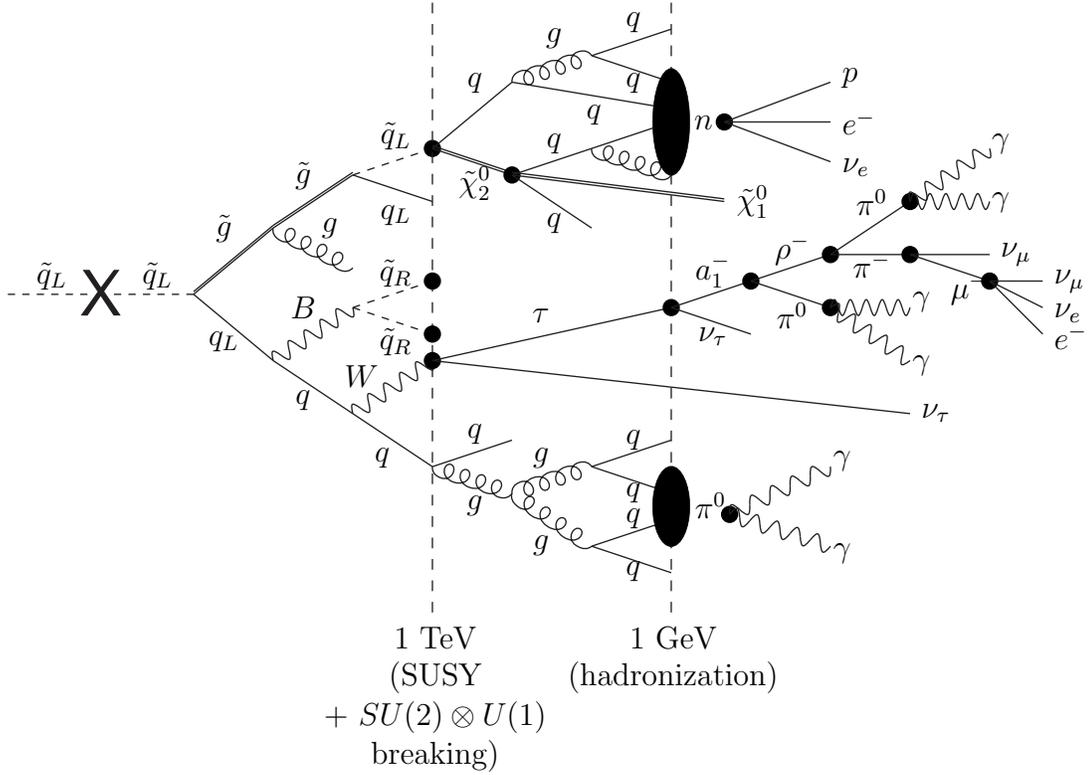
\end{center}
\vspace*{5mm}

\clearpage

\thispagestyle{empty}
\begin{figure}[h!]
\label{Organigram}
\setlength{\unitlength}{1cm}
\includegraphics{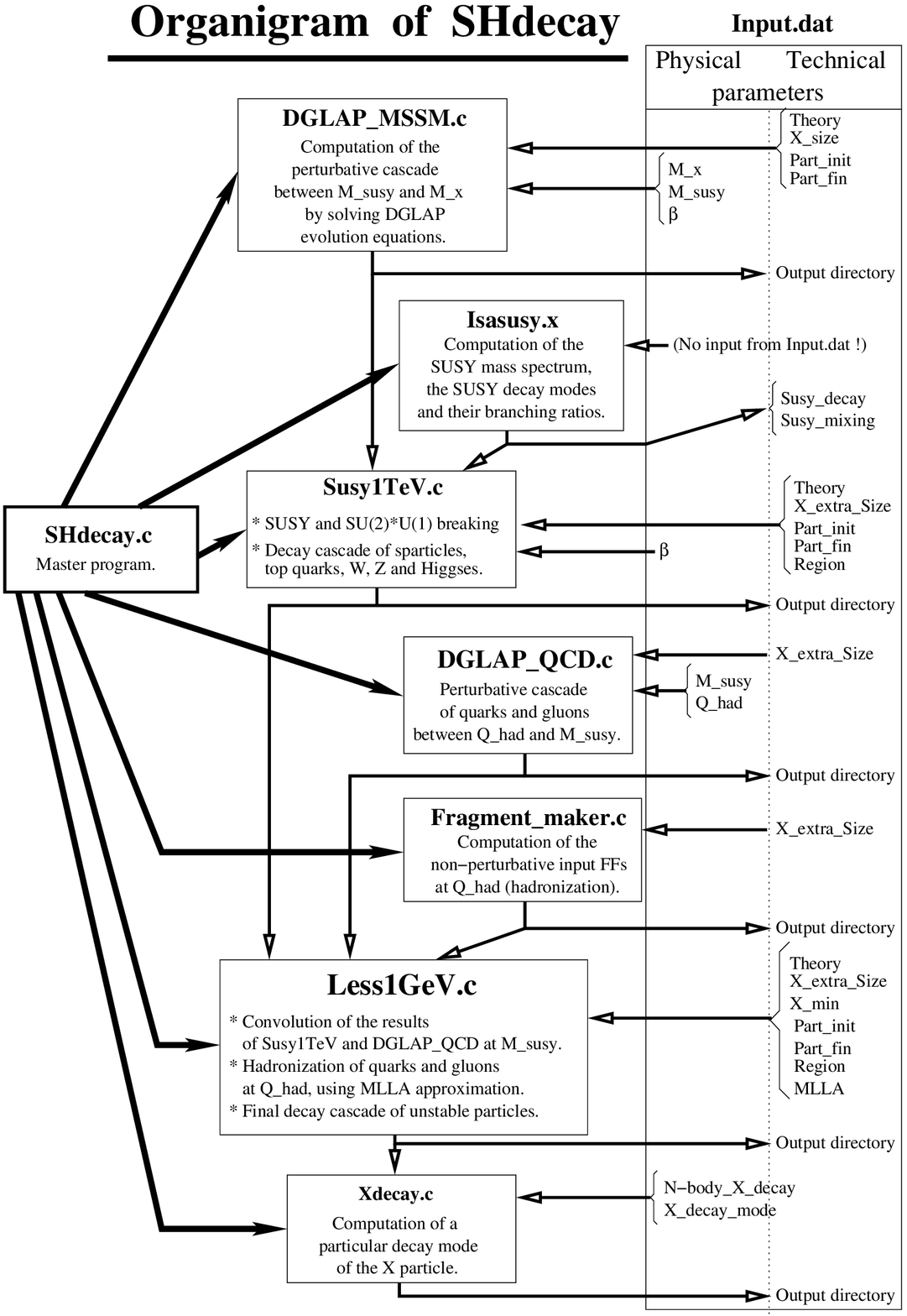}
\caption{Organigram of SHdecay with the interdependence between the
  different programs and the needed input parameters for each code.}
\end{figure}


\begin{thebibliography}{99}

\bibitem{reviewMartin} 
S.~P.~Martin, In Kane, G.L. (ed.): Perspectives
  on supersymmetry, 1-98, {\tt hep-ph/9709356}.

\bibitem{creat}
D.J.H. Chung, E.W. Kolb and A. Riotto, {\em Phys. Rev.} {\bf D60}
(1999) 0603504, {\tt hep-ph/9809453};
D.J. Chung, P. Crotty, E.W. Kolb and A. Riotto, {\em Phys. Rev.} {\bf D64}
(2001) 043503, {\tt hep-ph/0104100};
R. Allahverdi and M. Drees, {\em Phys. Rev. Lett.} {\bf 89} (2002) 091302,
{\tt hep-ph/0203118}, and {\em Phys. Rev.} {\bf D66} (2002) 063513,
{\tt hep-ph/0205246}.

\bibitem{reviewSigl}
P.~Bhattacharjee and G.~Sigl, {\em Phys. Rept.} {\bf 327} (2000) 109,
{\tt astro-ph/9811011}; L.~Anchordoqui, T.~Paul, S.~Reucroft and
  J.~Swain, Int.\ J.\ Mod.\ Phys.\ A {\bf 18} (2003) 2229 {\tt
    hep-ph/0206072}.

\bibitem{expts}
M.~A. Lawrence, R.~J.~O. Reid, and A.~A. Watson, {\em J. Phys.} {\bf G17}
(1991) 733;
D.~J. Bird {\em et al.,} {\em Astrophys. J.} {\bf 441} (1995) 144;
HiRes-MIA collab., T.~Abu-Zayyad {\em et al.,} {\em Astrophys. J.}
{\bf 557} (2001) 686, {\tt astro-ph/0010652};
AGASA collab., N.~Hayashida {\em et al.,} {\em Astrophys. J.} {\bf
522} (1999) 225, {\tt astro-ph/0008102}.

\bibitem{AP}
G.~Altarelli and G.~Parisi, {\em Nucl. Phys.} {\bf B126} (1977)
298.

\bibitem{bd1}
C.~Barbot and M.~Drees, {\em Phys. Lett.} {\bf B533} (2002) 107,
{\tt hep-ph/0202072}.

\bibitem{bd2} 
C. Barbot and M. Drees, {\tt hep-ph/0211406}, accepted
  for publication in Astroparticle Physics.

\bibitem{Berezinsky:2000}
V.~Berezinsky and M.~Kachelriess, {\em Phys. Rev.} {\bf D63} (2001) 034007,
{\tt hep-ph/0009053}.

\bibitem{Coriano:2001}
C.~Coriano and A.~E. Faraggi, {\em Phys. Rev.} {\bf D65} (2002)
075001, {\tt hep-ph/0106326}.

\bibitem{Sarkar:2001}
S.~Sarkar and R.~Toldra, {\em Nucl. Phys.} {\bf B621} (2002) 495, {\tt
hep-ph/0108098}.

\bibitem{ToldraLSP}
A.~Ibarra and R.~Toldra, {\em JHEP} {\bf 0206}, (2002) 006 {\tt
hep-ph/0202111}.


\bibitem{Isasusy}
H.~Baer, F.~E. Paige, S.~D. Protopopescu, and X.~Tata,
{\tt hep-ph/9305342}.

\bibitem{Poetter}
B.~A. Kniehl, G.~Kramer, and B.~Potter, {\em Nucl. Phys.} {\bf B582} (2000)
  514--536,
{\tt hep-ph/0010289}.

\bibitem{QCDrev}
E.~Reya {\em Phys. Rept.} {\bf 69} (1981)
195.

\bibitem{bdhh1}
C.~Barbot, M.~Drees, F.~Halzen, and D.~Hooper,
{\em Phys.\ Lett.} {\bf B555} (2003) 22,
{\tt hep-ph/0205230}.

\bibitem{bdhh2}
C.~Barbot, M.~Drees, F.~Halzen, and D.~Hooper,
{\tt hep-ph/0207133}, accepted for publication in Phys. Lett. B..

\bibitem{Jones}
S.~K. Jones and C.~H. Llewellyn~Smith, {\em Nucl. Phys.} {\bf B217} (1983)
145.

\bibitem{Basics_of_QCD}
Y.~L. Dokshitzer, V.~A. Khoze, A.~H. Mueller, and S.~I. Troian,, {\em Basics of
  perturbative QCD}.
\newblock Gif-sur-Yvette, France: Ed. Frontieres (1991) 274 p. (Basics of).

\bibitem{neutrinos}
P. Gondolo, G. Gelmini and S. Sarkar, {\em Nucl. Phys.} {\bf B392}
(1993) 111, {\tt hep-ph/9209236};
F. Halzen and D. Hooper, {\em Rept. Prog. Phys.} {\bf 65} (2002) 1025, 
{\tt astro-ph/0204527}.

\bibitem{LPHD}
Y.I. Azimov, Y.L. Dokshitzer, V.A. Khoze and S.I. Troian, {\em
Phys. Lett.} {\bf B165} (1985) 147, and {\em Z. Phys.} {\bf C27}
(1985) 65.

\bibitem{distorted_gaussian}
C.~P. Fong and B.~R. Webber, {\em Phys. Lett.} {\bf B229} (1989)
289.

\bibitem{limiting_spectrum}
  V. Berezinsky and M. Kachelriess, {\em Phys. Lett.} {\bf B434} (1998) 61, {\\t hep-ph/9803500}.
  

\end{thebibliography}
\end{document}